# ARTICLE

# Efficient data transport over multimode light-pipes with Megapixel images using differentiable ray tracing and Machine-learning

Joowon Lim [1]*, Jannes Gladrow [1]*, Douglas Kelly [1], Greg O'Shea [1], Govert Verkes [1†],

Ioan Stefanovici [1], Sebastian Nowozin [1†], and Benn Thomsen [1]

* Contributed equally to this work

† Work done at Microsoft Research

[1] Microsoft Research, 21 Station Road, Cambridge CB1 2FB, United Kingdom.

Corresponding author: Jannes Gladrow, jannes.gladrow@microsoft.com

# Abstract

Retrieving images transmitted through multi-mode fibers is of growing interest, thanks to their ability to confine and transport light efficiently in a compact system. Here, we demonstrate machine-learning-based decoding of large-scale digital images (pages), maximizing page capacity for optical storage applications. Using a millimeter-sized square cross-section waveguide, we image an 8-bit spatial light modulator, presenting data as a matrix of symbols. Normally, decoders will incur a prohibitive $O(n^2)$ computational scaling to decode *n* symbols in spatially scrambled data. However, by combining a digital twin of the setup with a U-Net, we can retrieve up to 66 kB using efficient convolutional operations only. We compare trainable ray-tracing-based with eigenmode-based twins and show the former to be superior thanks to its ability to overcome the simulation-to-experiment gap by adjusting to optical imperfections. We train the pipeline end-to-end using a differentiable mutual-information estimator based on the von-Mises distribution, generally applicable to phase-coding channels.

# INTRODUCTION

Transmitting optical signals over distances and retrieving them is a fundamental operation in many optical applications including imaging, communication systems[1,2], and optical storage which is the main motivation for this work. The size of transmittable images plays a crucial role for optical storage technologies such as holographic optical storage[3,4] which is seeing renewed interest driven by the increasing capacity and access rate requirements of cloud storage. In this application, to efficiently exploit the holographic media, it is necessary to support large page sizes (i.e., a large spatial number of spatial channels). Furthermore, data pages have to be transported compactly over relatively short distances within the optical storage device so that input and output optics, i.e., spatial light modulator (SLM) and camera, can be cost-effectively multiplexed across multiple zones in the media. Multi-mode fibers (MMF) have been explored as means for signal transmission in imaging[5] and digital communication[6], and augmented/virtual reality devices[7] as an alternative to bulky optical lens relays, due to their compactness and ability to confine and propagate light. In optical communication, the focus has been on longer transmission distances with a small number of spatial channels with the current record demonstrating transmission using 55 spatial channels (modes) over transmission distance of 29.5 km[8]. By contrast, in this work, we establish a different regime and show the transport of information over 315,000 spatial channels albeit for a much shorter distance of 75 mm through a square-cross-section 3.56 mm thick silica waveguide, referred to as light-pipe.

When images get transmitted through a MMF (or light-pipe), they experience heavy distortion in proportion to the length of the MMF and the dimension of the input image. In other words, it is extremely challenging to transmit a large image through a MMF and decode it from the measurement. This limits direct utilization of MMFs in imaging and communication applications. Recent progress has been made in undoing such distortion, generally following either a transmission-matrix (TM) approach[9–14] which relies on pre-measuring distortions, or statistical approaches, e.g., based on deep neural networks (DNNs)[15–18]. TM approaches attempt to estimate a matrix that maps the input to the output by transmitting orthogonal bases and measuring the corresponding outputs. By contrast, statistical approaches attempt to learn the distortion patterns by presenting various input and output pairs rather than relying explicitly on a TM to extract distortion features present in the dataset. Therefore, given a new input, statistical approaches interpolate or extrapolate learned features to predict the output signal.

In order to transmit $N$ independent signals using the TM approach, we need to construct a TM whose size increases quadratically, $N^2$, posing an inevitable bottleneck for larger number of signals. For example, well-established frequency-domain equalization (FDE) algorithms for multiple-input-multiple-output channels such as adaptive FDE rely on TMs leading to a $\mathcal{O}(NM\,log_2(M) + N^2M)$ runtime complexity where, $M$ is the channel memory (number of past data points that need to be taken into account) and $N$ is the number signals[19]. In other words, this highly efficient algorithm becomes quadratically slower in the number of in- and outputs, which is unproblematic in its applications such as long-haul fiber transmission where typical fiber mode numbers are low, on the order of tens of modes[20]. By contrast, in imaging applications higher mode numbers are typically required, straining the square runtime complexity of TM. The maximum effective image size demonstrated using TM, to our best knowledge, is 15,000[9] which we here surpass 21-fold. Beyond TMs, DNN-based approaches have shown promising results for restoring natural images, especially in the context of endoscopic applications[15–18]. However, such images are usually compressible due to their lower entropy. Trained DNNs used for input reconstruction may in this case memorize output features based on their internal weights rather than the input. However, Fan et al. pioneered the use of DNN to decode high-entropy data and demonstrated decoding of 20x20 1-bit-symbols in a data image (referred to henceforth as page)[21]. Here, *symbol* represents the spatial unit of data in a page presented on an SLM. Each symbol can present data with a certain bit precision. It is thus possible to train a DNN that learns the TM implicitly even for high-entropy data where every single input symbol matters[22]. However, as we show here, even this approach must necessarily suffer from the quadratic scaling of model sizes as the models need to learn the SLM-to-Camera pixel-to-pixel relationships and would thus be limited to small data pages. The aim in this work is to investigate the maximal transmittable page capacity through light-pipes for large images as a function of symbol number and light-pipe length. In order to revert the scrambling of information without building a TM or approximating it using a DNN, one has to reverse the spatial scrambling induced by the light-pipe using an efficient physics-based localizer. Once localized, we can feed the images into an efficient convolution-based model to decode large data pages without incurring a quadratic explosion of network parameters in the decoder model. We implement and compare differentiable mode-based and ray-based localizers. Importantly, the square-shape of our light-pipe simplifies the complexity involved in implementing the localizers: A circular shape would radially reflect rays emanating from the SLM plane resulting in smeared information over multiple pixels on the camera plane, by contrast, rays reflected by a square light-pipe converge to fewer points on the camera plane, and this simplifies the complexity involved in retrieving the original information as we show in the supplementary information. Even though the mode-based localizer can model light transmission through light-pipe more accurately in theory, we demonstrate that the ray-based localizer outperforms the mode-based localizer. This is because the differentiability of the ray-tracer allows it to adapt to deviations in the shape of the light-pipe which is unavoidable during manufacturing. In other words, the trainable ray-based twin can overcome the simulation-to-experiment gap, a well-known problem in other machine learning fields such as robotic control[23–25].

Finally, we describe a general framework to reason about the Shannon channel capacity of light-pipes which relies on an estimation of the mutual information (MI) between inputs and decoder outputs. MI is a measure between two random variables quantifying how much information is contained in one about the other. The estimated MI provides us with an upper bound to the number of bits that could possibly be transmitted. For each symbol location on a data page, the MI is estimated as a function of symbol size to study the ensuing trade-off between information per symbol (bit-depth) and the density of symbols (number of symbols presented on a page). A variational approach underpins our MI estimation[26]. The estimator is implemented as a convolutional Biternion neural-network[27] which is simultaneously trained to reconstruct the input symbol and predict its uncertainty. We therefore go beyond a standard mean-square loss lacking information about the uncertainty of prediction. Instead, we model the phase-encoded symbol as a gaussian distribution on the unit-circle with a mean value and a concentration (or inverse of variance) of the prediction of the mean. By estimating these values for each symbol in a maximum log-likelihood setting, we obtain a spatially resolved MI map that can be calculated directly from the uncertainty without having to apply statistical estimators over hundreds of large images[28]. Our approach is general and thus applicable to other multi-mode phase-coding channels.

## RESULTS

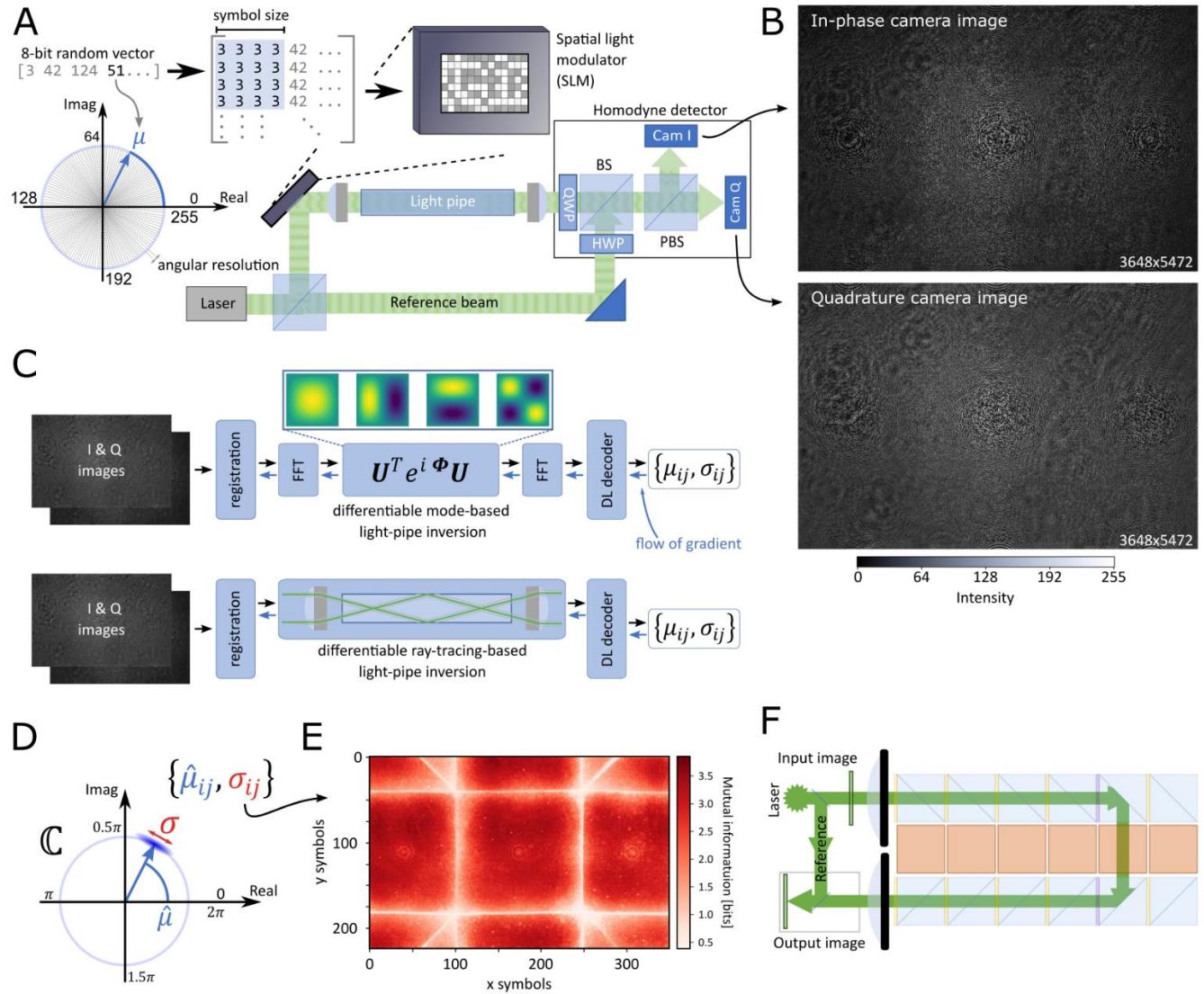

Figure 1. Main idea of the paper. **A** 8-bit symbols are displayed on an SLM over multiple pixels (decided by symbol size) to induce a phase delay of the signal beam. The signal beam modulated by the SLM is focused on the proximal facet, which is then followed by a light-pipe and another lens projecting to the homodyne detector. The sketch is not drawn to scale. QWP,

HWP, BS, and PBS stand for quarter-wave plate, half-wave plate, beam splitter, and polarizing beam splitter, respectively. **B** Examples of in-phase and quadrature camera images captured using the homodyne detector. In the central part of the image, a checkerboard symbol pattern is visible which is increasingly degraded towards the edges where phase distortion is stronger. **C** Schematic description of decoding pipeline which consists of a localizer followed by a convolution-based decoder. The localizer can be either mode-based or ray-based. The localizers recover locality which is lost during light-pipe propagation. **D** The output of the decoder predicts the original phase value along with the uncertainty of the prediction. **E** Spatial total MI distribution retrieved from the uncertainty map for a symbol size of 4x4 pixels. **F** Sketch of the light-pipe data page routing concept. A sequence of cubic switches (light blue, e.g., polarization-based) form a light-pipe and route data pages to and from the relevant locations inside an optical storage medium (orange). The reference beam for the local oscillator of the homodyne detector can be routed outside of the light-pipe and media. It can be transmitted in free space as a collimated beam.

## *Main idea*

In digital systems such as data storage, binary messages are typically encoded in terms of symbols which can take values $\{0, ..., 2^b - 1\}$ where b is the bit-depth of the message. Each symbol is presented over a certain number of SLM pixels (see Fig. 1A). For a fixed total number of SLM pixels, the number of symbols therefore increases quadratically with decreasing symbol size. However, given the presence of noise in experiment, the smaller the symbol size (in pixels), the lower the signal-to-noise ratio, constraining the decodable bit-depth for each symbol. We utilized the full 8-bit depth of our SLM, that is, we set $b = 8$ (see methods), to represent phase-encoded data images, which are therefore 8-bit random matrices. Each symbol is represented over multiple SLM pixels from 16x16 to 2x2 SLM pixels, which is referred to as the symbol size. The total amount of displayable information in a single image is thus $N_y \times N_x \times 8$ bit ($N_y$ and $N_x$ represent the number of symbols not the number of pixels), and ranges from 4.8 kB to 2.5 MB as a function of symbol size. However, only a fraction of these theoretical data pages will be retrievable due to noise, quantization by cameras, and the scrambling induced when transmitting data pages through light-pipes.

We consider millimeter-sized light-pipes consisting of a silica rod core and air cladding. The square cross-section was chosen to be congruent with the shape of SLM pixels which we aim to image using the light-pipe in our setup (see Fig. 1A). In addition, unlike circular cross-section waveguides where rays are radially spread over multiple locations once reflected inside the waveguide, the square cross-section minimizes the number of points where they converge to. This is advantageous as it simplifies sampling requirements for ray-based localizer. Please see the supplementary information for detailed information about the comparison between square and circular cross-sections. We propagate phase-encoded SLM images through a 2f-launch lens into a 75mm-long light-pipe and, using another 2f-de-launch lens, onto a two-camera homodyne detector as shown in Fig. 1A.

We transmit up to 315,000 symbols and capture them oversampled on megapixel cameras (see Fig. 1B). The homodyne detector simultaneously captures two images as in-phase and quadrature (see Fig. 1B) where each camera detects the intensity of coherently summed signal and reference beams. The signal beam contribution to the quadrature image is a $\pi/2$-delayed version of that of the in-phase image such that both images combined give access to the full electric field. In previous works on imaging through MMF, detection was either intensity-only[29], which is inherently lossy, or based on off-axis holography[12,30]. In off-axis holography, one needs to ensure that 0-order and ±1-order signals to not overlap, which comes at the cost of limiting the spatial-bandwidth product[31]. In this work, we introduce an adaptation of the homodyne detector described in [32] using a dual camera system instead of a switchable retarder to measure the full optical field at the distal facet without limiting spatial- or temporal-bandwidth. Please refer to the supplementary information for detailed explanation about in-phase and quadrature images.

When we try to decode original symbols from the camera measurements on larger images, it is infeasible to invert the distortion process solely using a convolutional DNN because of the quadratic scaling of parameters. To be specific, due to the information scrambling by the light-pipe, we need a DNN whose receptive field is large enough to capture the information delocalization (we explain this in more detail in the next subsection). Within a large receptive field, a DNN should learn the pixel-to-pixel relation to reverse the delocalization, which requires $O(n^2)$ all-to-all connections if no further inductive biases are used. To address this shortcoming, we introduce physics-based machine-learning modules which pre-compensate for the information scrambling. The output of these specialized modules is then fed into a convolutional DNN. The overall decoder system is thus a hybrid of physics-inspired pre-compensation modules and neural networks (see Fig. 1C). We introduce two types of pre-compensation modules. First, we implement a mode-based approach using analytical descriptions of mode propagation for light-pipes. However, the mode-based approach cannot handle deviations of the light-pipe from square cross-sections, which we found to be unavoidable due to manufacturing tolerance. We therefore propose a second type of pre-compensation module, based on differentiable ray-tracing. This module is flexible enough to implement real-world light-pipe shapes and, crucially, enables us to adjust automatically to slight deviations of our light-pipe from a perfect square cross-section and even incorporate accurate models of the lenses we use in our setup. As a result, the ray-based model shows superior performance on experimental data than the mode-based approach. Both mode-based and ray-based localizers are differentiable which is critical as it allows us to jointly train further modules downstream from the pre-compensation modules. Specifically, our dual-camera detector requires image registration, which here is also learned, driven by gradients passed down through the specialized modules from the decoder.

When camera images are processed by a light-weight decoder system consisting of a digital twin of the light-pipe and lenses as well as a convolutional neural network (CNN), the role of the CNN is to retrieve the phase modulation at the SLM as a continuous angle and estimate the uncertainty of this prediction (see Fig. 1D). More precisely, instead of using mean-square or cross-entropy losses, our model is trained to predict parameters of a von-Mises distribution $p(\vartheta|\mu,\kappa) = \frac{\exp(\kappa\cos(\vartheta-\mu))}{2\pi I_0(\kappa)}$ on the unit-circle where $\vartheta$ is the true phase angle that we send to the SLM, $\mu$ is the expected value of the phase angle, and $\kappa$ denotes the concentration (inverse variance, i.e. $\kappa_{ij} = \sigma_{ij}^{-2}$) of the distribution. As the angle values are in the argument of the cosine, we do not have to unwrap the phase values separately. For a pair of in-phase and quadrature input images, the network will thus output not only mean values of each symbol, $\mu_{ij}$, but also the concentration (or inverse of uncertainty) of the prediction, $\kappa_{ij}$ (see Fig. 1C-1D). Please see the supplementary information for details of the implementation. Training is achieved by minimizing the negative log-likelihood of the true angle $\vartheta$. Predicting a full distribution of the phase angle rather than just the expected value enables us to make a spatially resolved MI estimate for each symbol. The MI (in bits) between symbol angles displayed on the SLM ($\vartheta$) and angles predicted by the network ($\mu$), can be written as

$$I_\kappa(\vartheta; \mu) = H(\vartheta) - H_\kappa(\vartheta|\mu) \approx \log_2(2\pi) + \int_0^{2\pi} p(\vartheta'|\mu,\kappa) \log_2(p(\vartheta'|\mu,\kappa)) d\vartheta' = -\log_2(I_0(\kappa)) + \kappa \frac{I_0(\kappa)}{\log(2) I_1(\kappa)},$$

where $H$ denotes the entropies of the respective variables, while $I_0$ and $I_1$ denote the modified Bessel function of the first kind of order zero and one. The equation above assumes that all symbol angles are equally distributed, which is approximately the case (see Fig. 1A). During training, the MI increases driven by decreasing network uncertainty. In Fig. 1E, we present one example of our results (4x4 symbol size) showing the estimated spatial distribution of MI. Such image transport and analysis of MI on the transmitted image are essential in applications where we transport and decode high density data such as holographic optical storage[3,4]. For this technology to be scalable and competitive, it is crucial to guide data-encoded optical images to multiple target locations in the media using compact optical system, to fully utilize its capacity. We can use light-pipes to route data pages to target locations albeit in distorted form, inside optical storage media as sketched in Fig. 1F.

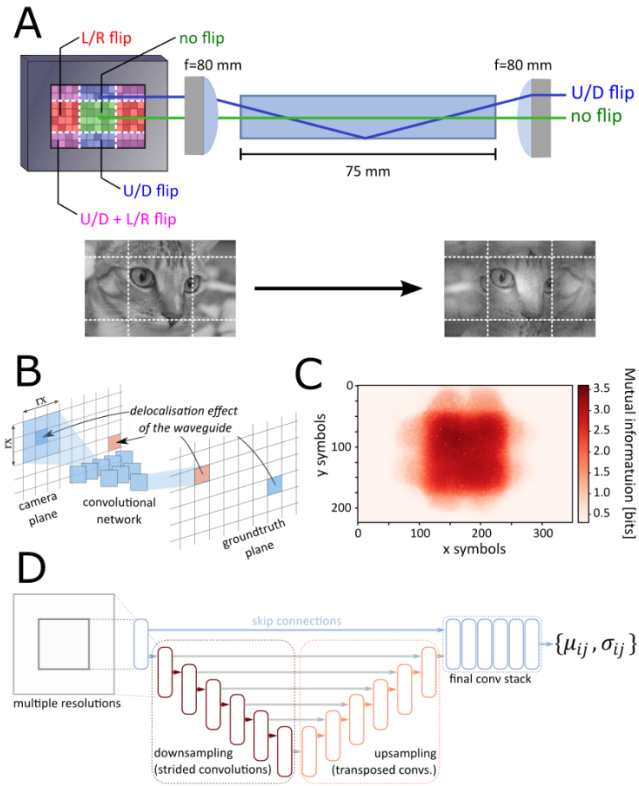

Figure 2. Delocalization of information and the receptive field of the CNN-based decoder without localizer. **A** Schematic description of delocalization within the light-pipe. Depending on the location on SLM, each pixel experience different numbers of total internal reflection. L/R means left/right, and U/D means up/down. The transformation is applied to Chelsea cat image as an example. The lens-induced global flip is corrected. **B** Simplified sketch of the scrambling-decode problem: A convolutional network will read in a receptive field around the symbol from the input (camera images) to predict the symbol value at this position (output). However, since the inputs are scrambled spatially, the network will likely predict the value of a different symbol (red instead of blue). **C** MI map obtained with a U-Net decoder without localizer. Only the central unscrambled zone is decodable. **D** Sketch of the CNN (U-Net) architecture that we use as signal decoder. The sketch highlights the fact that the U-Net processes the incoming images at various resolutions corresponding to different receptive field sizes (only two highest resolutions are shown), while moving the convolved signal forward via skip-connections. Blue boxes are 2D convolutions with stride 1, while red boxes have strides of 3, and orange boxes indicate 2D transposed convolutions also with stride 3. The details of the U-Net including kernel sizes and numbers are given in the supplementary in Fig. S3.

### *Information retrieval study varying symbol sizes using differentiable localizers*

In Fig. 2A, we explain how delocalization occurs within light-pipes using ray optics in detail. Rays emanating from the central zone of SLM propagate without any reflection. By contrast, rays coming from the up/down zones (blue ray) get reflected once. As a result, these zones appear flipped on the camera plane. The same happens along the horizontal direction. Therefore, when we transmit an image (cat image in Fig. 2A) displayed on SLM through our optical system, the central part of the original image will remain localized but other parts will flip around various axes. A CNN will read in a receptive field around the symbol to predict the symbol value at this position (output) from the input (camera images). However, since the inputs are scrambled (relative to the ground truth), without an inverting digital twin, the decoder is only able to estimate the symbol phase-profile at the central zone as shown in Fig. 2B. In other words, the light-pipe delocalizes parts of the input image over an area that exceeds the receptive field of the CNN (see Fig. 2C). A schematic description of the CNN we use (a U-Net) is presented in Fig. 2D. Please refer to methods and supplementary information for details. Algorithmically, the question of which symbol appears in which pixel is of quadratic order in the number of symbols (since we need to consider all possible combinations between symbols). In our case, this is a prohibitively large number (315,000 symbols at maximum). A typical approach to this problem is to measure and build a fully-connected TM to map the electric field from the SLM plane to the camera plane[9–12]. To escape the quadratic scaling that this would incur, we use a square-cross-section light-pipe on the experimental side minimizing spatial symbol spread. On the digital side, we implement digital twins to obviate the need for all-to-all layers in a deep-learning decoder. The main role of the digital twin is thus to restore locality to the image such that the network is given the information it needs to retrieve the phase profile on the SLM plane for each symbol. We hence refer to our digital twins also as localizers.

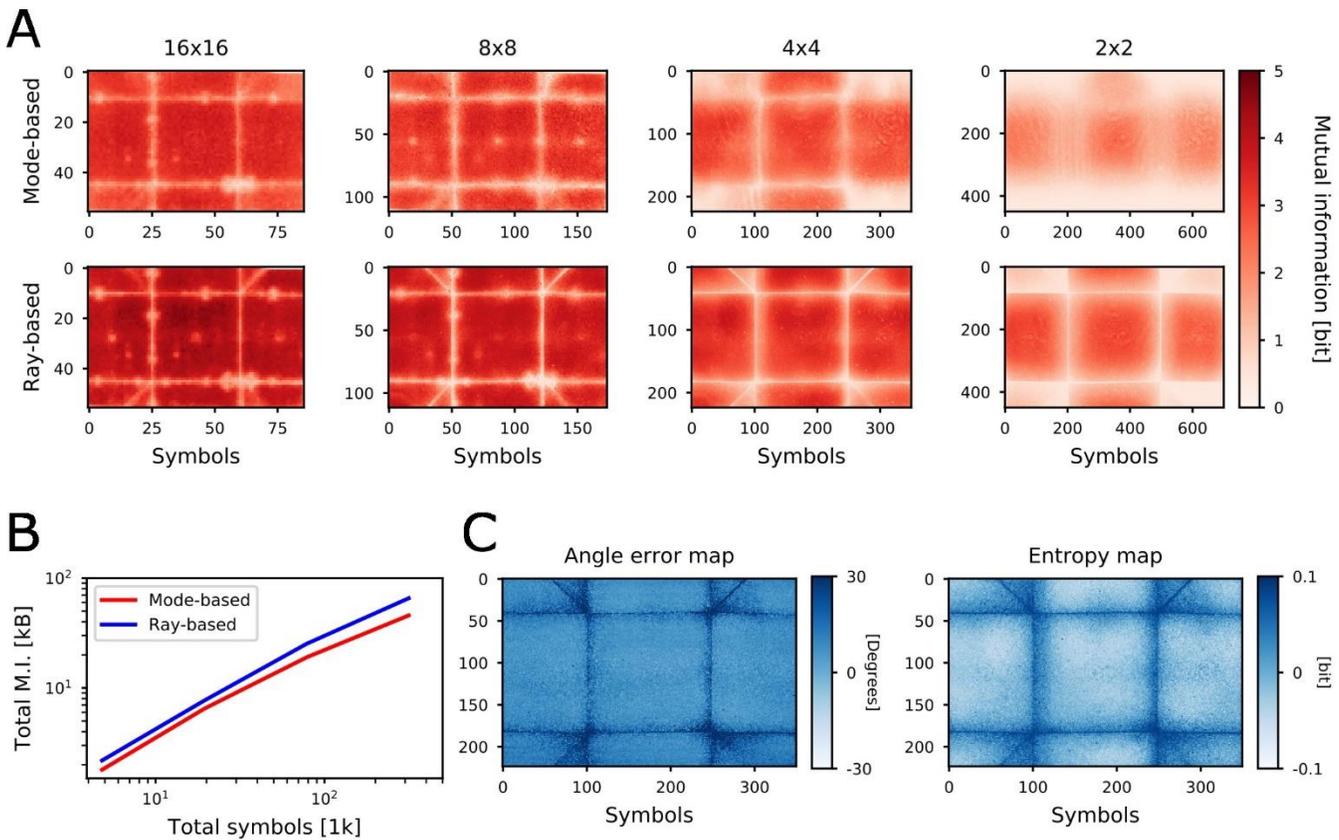

Figure 3. Total MI comparisons of experimental data with varying symbol size. A The first and second rows show the MI maps based on the mode-based and ray-based localizers, respectively. From left to right columns, the symbol size varies from 16x16 to 2x2 pixels. B Total MI for different symbol sizes shown in A. Total symbols denote the total number of symbols present in one page which increases with the decrease of symbol size. C A comparison between the estimated uncertainty from the decoder with the calculated (differential) entropy map from the test dataset.

By nature, the mode-based and ray-based localizers act in different ways: the mode-based localizer, by virtue of being a wave-based model, is a richer representation of light-pipe propagation. It inverts the spatial delocalization as well as the modal phase dispersion and hence needs to be run at both test and training time. The runtime complexity of the mode-based localizer is dominated by the matrix multiplications in Fig. 1C, $\boldsymbol{U}^T e^{i\boldsymbol{\Phi}} \boldsymbol{U}$. Assuming $N$ symbols are transmitted in a square-shaped signal $[\sqrt{N}, \sqrt{N}]$ for simplicity, these matrices have shapes $\boldsymbol{U}, \boldsymbol{\Phi} \in \mathbb{R}^{\sqrt{N} \times \sqrt{N}}$, such that the runtime amounts to $\mathcal{O}(N^{3/2})$, better than the usual $\mathcal{O}(N^2)$ complexity. By contrast, our ray-based localizer only simulates the geometric aspects of our optical transmission pipeline. Therefore, the ray-based localizer outputs a grid of points on the camera indicating where rays emanating from symbols on the SLM impinge on the camera(s). This grid can then be used to reassemble the image in such a way that restores locality. For the ray-based localizer, inversion of the phase response is thus absorbed into the deep-learning decoder. Counter-intuitively, the ray-based pipeline is computationally light since we can trace multiple rays in parallel, independently, and ray-tracing operations only need to be carried out during training. At test time, the grid is static and can simply be loaded. The ray-tracing localizer has therefore $\mathcal{O}(N)$ runtime complexity in the number of symbols, improving on the mode-based runtime complexity. Since our U-Net decoder is convolutional, its runtime is also linear in the number of symbols, the overall complexity remains linear. One would expect the mode-based-decoder pipeline to perform better than the ray-based approach given its physical accuracy. However, we show below that the opposite is the case for experimental data.

In Fig. 3A, we vary the symbol sizes from 16x16 to 2x2 pixels and display the total MI maps for both mode-based and ray-based localizers. A symbol size of 1x1 pixel results in 1.3M symbols leading to a low signal-to-noise ratio, too low for the camera registration module to converge in the experimental system. We therefore omit this datapoint here. In Fig. 3A, we show how much information can be encoded in different regions of the SLM in terms of MI. The corresponding total MI values are plotted in Fig. 3B, confirming that the ray-based localizer outperforms the mode-based localizer in all cases. For instance, for symbol sizes of 4x4 and 2x2 SLM pixels, we obtain usable data page sizes of 19 kB and 46 kB using the mode-based localizer, and 25 kB and 66 kB using the ray-based localizer. In other words, we achieved 31% (4x4 pixels) and 43% (2x2 pixels) better performances in decoding using the ray-based localizer. Finally, to confirm that our network faithfully predicts the uncertainty of its own phase prediction, we averaged the angle error over images in the test dataset and compared that to the averaged entropy map. A comparison in Fig. 3C shows consistency between the two images.

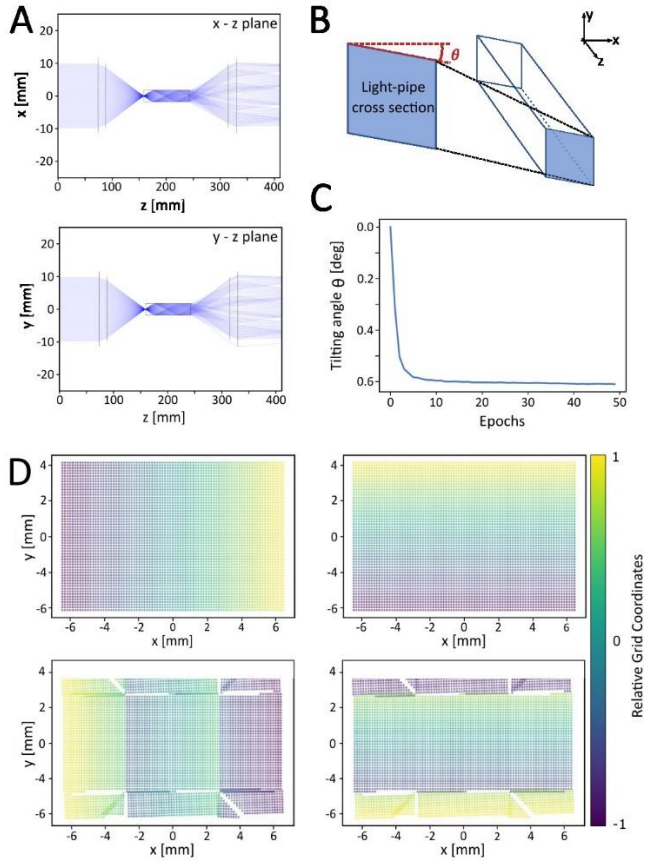

Figure 4 Differentiable ray tracing. **A** Plot of rays propagating through lenses and light-pipe. Each lens is a thick lens that consists of two materials (three surfaces). **B** Schematic description of the parallelogram angle. **C** The fitted parallelogram angle during training. **D** The top row displays the positions of rays on the SLM plane and they are color-coded in $x$ (left) and $y$ (right) coordinate values, and the bottom row displays the positions of rays on the camera plane after propagating through a parallelogram light-pipe. The lens-induced global flip is corrected.

This is counterintuitive as the mode-based localizer is more accurate and able to capture the wave nature of light better than the ray-based localizer. However, unlike the mode-based localizer, the ray-based localizer is able to accommodate deviations of our experimental light-pipe from its designated shape resulting in increased performance for experimental light-pipes. We discovered that the light-pipe deviates by a fraction of a degree from a perfect square shape and resembles a parallelogram, causing the modal shapes to differ from those of a perfect (square cross-section) light-pipe (Please refer to the supplementary information for the derivation.). In the first row of Fig. 3A, we can see the gradual degradation of total MI in the diagonal zones and, for small symbol sizes (4x4 and below), also in the top and bottom zones. We address this problem in the ray-tracing picture, where it is relatively straightforward to relax the light-pipe shape into a parallelogram (see methods). Figure 4A shows ray propagation through our digital twin (a light-pipe along with two lenses). Moreover, since our ray-tracer is fully differentiable, i.e., it is able to apply the chain rule to lens or light-pipe parameters, we are able to 'fit' the parallelogram angle during training of the full decoder system. The parallelogram angle converges to around -0.6° (see Fig. 4B-C) starting from an initial value of 0°. This is an example of inverse ray tracing[33–35] where we can recover aspects of the geometry of our setup from acquired images without measuring them directly. In Fig. 4D, the top row displays the positions of rays on the SLM plane, color-coded in $x$ an $y$ coordinate values. The bottom row shows the ray positions on the camera plane after propagating through the parallelogram light-pipe. We can clearly see the overlap of rays in the diagonal zones, which explains why it is challenging to recover information in these diagonal zones particularly using the mode-based localizer. We use the ray-positions to resample camera images to restore locality. As a result, in Fig. 3A. We can see that the ray-based localizer can decode diagonal zones better than the mode-based localizer.

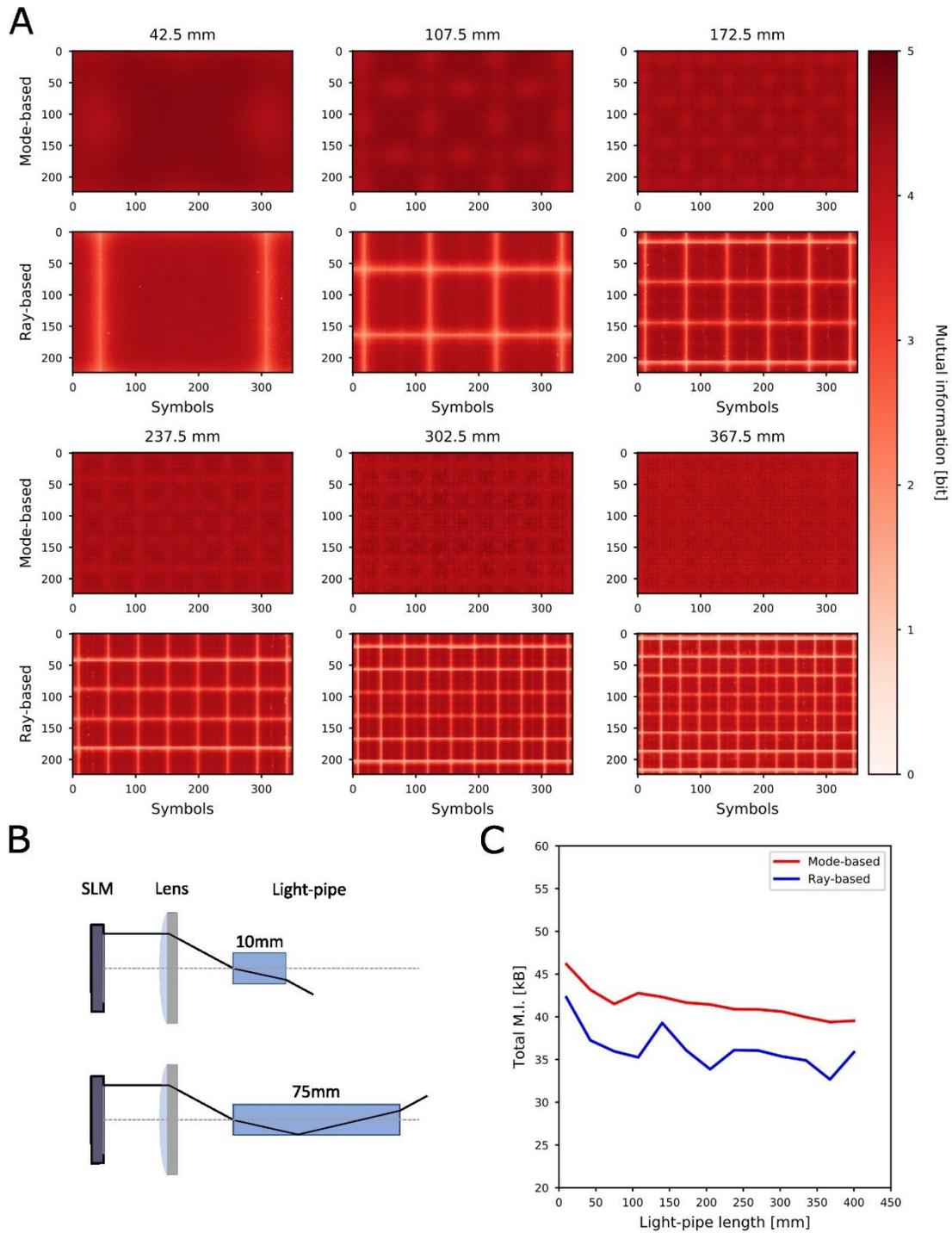

Figure 5. Results of MI estimation on simulation data for various light-pipe lengths. **A** Total MI maps for different light-pipe lengths using mode-based (the first and third rows) or ray-based (the second and fourth rows) localizers. **B** Schematic description of the increase in total internal reflection number with the increase of light-pipe length. **C** Quantitative evaluation of total MI per image with the length of light-pipe for two localizers.

*Information retrieval study varying light-pipe lengths*

To study the mechanism by which we lose information, we compare the performance of our localizers for different light-pipe lengths on synthetic data generated using simulations. It includes a model of our homodyne detector system, lenses as well as a perfect (square cross-section) light-pipe (see methods). We vary the length of the light-pipe from 10 mm to 400 mm at an interval of 32.5 mm (In Fig. 5A, we show only subsets of them.). The 32.5-mm interval was chosen to increase the number of total internal reflections each time as shown in Fig. 5B. Due to the perfect square cross-section in simulations, the mode-based localizer outperforms the ray-based localizer as we use the same forward and inverse model for the light propagation as shown in Fig. 5C. For example, for the case of 75 mm, the mode-based localizer outperforms the ray-based localizer by 15%. Interestingly, in both cases, we observe the emergence of additional low MI regions (in-between squares) with increasing light-pipe length in Fig. 5A (compared to Fig. 3A). These are areas where locality jumps from one zone to another. In other words, local and non-local information on SLM is coherently summed, resulting in lowered intensity. In these areas several symbols overlap, making it harder for any decoder model to distinguish individual symbol values. The signal in these zones is thus more susceptible to noise resulting in lowered MI. Therefore, even for the mode-based localizer, we can see a decrease of total MI over increasing light-pipe length, and a general mechanism of information loss. The ray-based localizer shares the same tendency, but it has a performance penalty compared to the mode-based localizer over increasing light-pipe length. In our ray-based model, we send only one ray per pixel, such that we cannot fully retrieve the electric field of symbols that spread over two or more locations. As a result, the ray-based localizer is more susceptible to the emergence of these regions. Since the camera sensor is rectangular, such regions appear along the horizontal direction first (Fig. 5A, 10 mm). With increasing light-pipe length, there are discrete increments of these low MI regions which render the MI decrease non-monotonic. Nevertheless, in the previously discussed experimental case, the ray-based localizer outperforms the mode-based localizer as it can auto-adjust to the parallelogram-cross-section effectively.

## DISCUSSION

In this paper, we propose and demonstrate a framework for transmitting maximal-entropy images (data) through a light-pipe. Given that data is encoded in the phase of light, we utilize a homodyne detector to fully capture the phase information without limiting the spatial bandwidth. We can deliver up to 66 kB of information through the light-pipe without relying on auxiliary imaging systems. The rate of transmittable data pages is limited by the update rate of the SLM. With currently achievable SLM update rates of around 1.5 kHz[36], this would translate to 100 MB/s if we indeed transmitted up to the MI estimated here. Whilst this is lower than the maximum sequential transfer rate of competing technologies such as hard-disk drives (HDDs) at 285 MB/s[37], it is much higher than the random read access rate of such an HDD which delivers 168 Input/output operations per second giving a throughput of 11 MB/s for the same page size[37]. Our transmission rates are, therefore, of the right magnitude even if the effectively transmitted information per page ends up lower.

One critical challenge in retrieving the original signal lies in the fact that the locality of original information is lost while light propagates through the light-pipe. We show that a convolutional decoder on its own can only decode the central region where the locality is preserved. Networks with all-to-all connectivity, such as transformer type networks[38] are infeasible here due to their quadratic computational scaling with the number of symbols.

Therefore, we implement two localizers which exploit wave (mode-based) and ray (ray-based) properties of light. In theory, the mode-based localizer should be superior to the ray-based localizer as we demonstrated using the simulation. For 4x4 SLM pixels with the light-pipe length of 75 mm, the mode-based localizer outperformed the ray-based localizer by 15%. The ray-based model is generally flexible while the mode-based model requires analytical descriptions of eigenmodes of specific geometric shapes such as square or cylinder. We implemented our ray-model to be differentiable so any gradient propagated from the loss function to the output of ray-tracer can be propagated through the ray-tracer whose parameters can then adjust to the experimental configuration. Therefore, we can learn the parallelogram angle of the light-pipe which we find unavoidable during manufacturing. We show that, by recovering this aspect of the experimental set-up, the ray-based localizer performs better than the mode-based localizer. For 4x4 SLM pixels with the light-pipe length of 75 mm, the ray-based localizer outperformed the mode-based localizer by 31% on experimental data. This ability of the localization-pipeline to adapt to and uncover optical geometry is, we believe, widely applicable in waveguide image transmission. However, the ray-based model also has its limitations as shown in the simulation study. In Fig. 5, we can see the limitations of the current ray-based model, resulting from its inability to reconstruct symbols that scatter to multiple locations on the camera.

While our light-pipe is rigid so it cannot bend or twist like conventional MMFs, thermal conversion or mechanical vibration can lead to a drift in the signal over time. In order to estimate the drift in our setup, we conducted an additional experiment to characterize the stability of our set up over 24 hours and found the maximum drift after warm-up phase to be within a single camera pixel in average. Details can be found in the supplementary materials. Nevertheless, general care should be taken to isolate the set-up thermally and mechanically to minimize the need for retraining our decoder networks[18].

Lastly, throughout the paper, we trained a DNN to predict the parameters of von-Mises distribution, which can be used to calculate spatially resolved MI distributions. As MI is a measure between two random variables, quantifying how much information is contained in one about the other, it provides an upper bound to the number of bits that could possibly be transmitted. This framework is thus general and can be applied to any information transportation system beyond what has been demonstrated in this work. In accord with theory, the MI only upper bounds the retrievable data per symbol. To use our channel as a communication channel, we need to retrieve symbols, that is, exchange the angle regression in our model for a symbol classification. In summary, we believe that our work opens a new route for maximal-information transmission through light-pipes and MMFs, along with a mechanism to self-calibrate during training. Such image transport is essential in imaging[1] or communication[2] applications such as holographic optical data storage[3,4]. Furthermore, the ray-based framework could also be used to design optical hardware in conjunction with deep neural networks.

## MATERIALS AND METHODS

### *Experimental Setup*

532 nm light from a single longitudinal mode source is expanded and split between "signal" and "reference" paths by a polarizing beam splitter (PBS), with the ratio of power between each path controllable by adjusting the angle of a half wave plate. The reference beam is directed to the reference port of the homodyne detector system via a half wave plate to realign the reference polarization with the signal arm. The signal beam is reflected by a 50:50 beam splitter onto a nematic Liquid crystal on silicon SLM (Meadowlark HSP1920-488-800) comprising a 1920×1152 array of 9.2 µm pixels capable of imposing $2\pi$ phase delay at 532 nm with 8 bits of precision. The reflected, encoded beam is transmitted by the beam splitter and is focused onto the input facet of the light-pipe by a f=80 mm doublet (AC508-080-A-ML, Thorlabs). The output of the light-pipe (UV fused silica without cladding, input facet side=3.56 mm, length=75 mm, OPCO Laboratory Inc.) is collected by a second f=80 mm doublet and enters the signal port of the homodyne detector. In the supplementary material, we specify the conditions that light-pipe should meet in order to fully transport the information. The signal is converted to circular polarization by a quarter wave plate with fast axis oriented at 45° to the signal polarization. This introduces the required $\pi/2$ phase shift between the two orthogonal polarization components of the signal before combining with the reference in the 50:50 beam splitter. The half waveplate in the reference arm is used to adjust the power of the reference between the two orthogonal polarization states. Finally, the in-phase and quadrature components are split to two output ports by a PBS. At each output port is a camera (FLIR BFS-U3-200S6M-C) comprising an imaging chip of 5472×3648 2.4 µm pixels. Please refer to the supplementary information for a detailed explanation of the principles of the homodyne detector system.

### *Decoder system:*

Our entire decoder pipeline, including the mode-based and ray-based localizers, was implemented in Pytorch (1.6)[39]. In the following, we give a summary of each component of this system.

- **Mode-based localizer:**

The propagation of TE-modes in a light-pipe is a classic problem. The propagator of the electrical field can be obtained from solving Helmholtz' Equation $\Delta E(x,y,z) + \varepsilon_r(x,y,z)k^2 E(x,y,z) = 0$ where $E(x,y,z)$ is the electrical field, and $\varepsilon_r(x,y,z)$ and $k$ are the relative electric permittivity value (which is square of refractive index value) and free space wave-number respectively. The equation comes with boundary conditions; here we assume absorbing boundaries as an approximation, i.e., $E(x,y,z) = 0$ on the boundaries. Then, we can represent the modes with sinusoidal bases with different mode propagation speeds. For the details, please refer to the supplementary information.

- **Ray-based localizer:**

Differentiable ray tracing has recently attracted a lot of attention[33–35]. Here, we implement a differentiable source-to-camera ray-tracer. Fundamentally, many aspects of ray tracing are differentiable: For instance, when a ray gets reflected by a partially absorbing surface, the amplitude of the ray is reduced by a factor, a linear operation that allows gradients to be computed with respect to this material parameter. Likewise, the position variable of rays can carry gradients with respect to the geometry of the optical setup. In our case, the most important variable is the parallelogram angle of our light-pipe as discussed in the results section. For further details, please refer to the supplementary information.

- **Decoder U-Net architecture in detail:**

Our system captures in-phase and quadrature camera images, from which we sample to generate input images to the decoder. When transmitting $(N_x, N_y)$ symbols (not pixels), we display one symbol over multiple pixels (decided by symbol size). The pixel size of the camera is smaller than that of the SLM. When we sample camera images to generate input images for the decoder pipeline, the sampled image size is $(N_{sampling} N_x, N_{sampling} N_y)$, where $N_{sampling}$ represents the oversampling rate. For experimental data, we use different $N_{sampling}$ values depending on symbol size ($N_{symbol}$): $N_{sampling} = 4$ (for $N_{symbol} = 2$ and $N_{symbol} = 4$), $N_{sampling} = 8$ (for $N_{symbol} = 8$), $N_{sampling} = 16$ (for $N_{symbol} = 16$). For simulation data, we use $N_{sampling} = 4$ (for $N_{symbol} = 4$). Please refer to the supplementary information for details.

The input to the decoder consists of 4 channel images: two sampled camera images and two positional encoding images ($x$ and $y$ grid images). We use a U-Net type architecture that consists of down-sampling blocks followed by up-sampling blocks and a few convolutional layers[40]. The architecture is sketched in Fig. 2D, and the detailed structure is summarized in the supplementary information. After the up-sampling blocks, the image size is the same as the size of the input, and we need to further down-sample to match the number of symbols. In other words, as explained in the previous section, the image size after the up-sampling block is $(N_{sampling} N_x, N_{sampling} N_y)$, therefore, it is followed by additional convolution layers which perform down sampling to $(N_x, N_y)$, and we control the strides of the convolution layers depending on $N_{sampling}$. The details can be found in the supplementary information.

- **Training:**

For training, we use pre-recorded datasets from our setup or our simulation. For each page, we save the ground truth symbol value (8-bit integer) and the ensuing image. We can vary the symbol size, which determines the total symbol number. The total page numbers are 500 for all datasets. Training, validation, and test datasets are split with the ratio of 0.9, 0.05, and 0.05. The other training parameters and procedures are described in the supplementary information.

- **Compute resources:**

Our machine-learning pipeline was trained with a batch size of 2 end-to-end on an Azure NC24 virtual machine with one Nvidia K80 graphics processing unit with 24 GB of memory using Azure's AML service.

**ACKNOWLEDGMENTS**

We want to thank Sarah Lewis for discussions and her ongoing interest in the project. We also wish to thank Thomas Karagiannis, Kai Shi, Babak Rahmani, Erika Aranas, as well as Bruno Magalhaes for fruitful discussions. Finally, we are deeply indebted to Hannes Schulz and the Amulet team for helping us set up our cloud-based machine learning training and testing pipeline.

**CONFLICT OF INTERESTS**

The authors declare that they have no conflict of interest.

**CONTRIBUTIONS**

J.L. wrote the text, carried out most of the computations and simulations and contributed to the code base, J.G. guided the research, carried out experiments, contributed to the text, and authored the code base. D.K. built and aligned the optical setup and contributed to figures. G.OS. developed the software for addressing the hardware and experiment automation. G.V. and I.S. contributed to the information-theoretic metric, I.S. further contributed to designing the decoder model. B.T. had the initial idea and supervised the project.

**REFERENCES**

1. Flusberg, B. A. *et al.* Fiber-optic fluorescence imaging. *Nat Methods* **2**, 941–950 (2005).

2. Agrawal, G. P. *Fiber-optic communication systems*. (John Wiley & Sons, 2012).

3. Coufal, H. J., Psaltis, D. & Sincerbox, G. T. *Holographic data storage*. vol. 8 (Springer, 2000).

4. Chatzieleftheriou, A., Stefanovici, I., Narayanan, D., Thomsen, B. & Rowstron, A. Could cloud storage be disrupted in the next decade? in *12th USENIX Workshop on Hot Topics in Storage and File Systems (HotStorage 20)* (2020).

5. Papadopoulos, I. N., Farahi, S., Moser, C. & Psaltis, D. Focusing and scanning light through a multimode optical fiber using digital phase conjugation. *Opt Express* **20**, 10583–10590 (2012).

6. Richardson, D. J., Fini, J. M. & Nelson, L. E. Space-division multiplexing in optical fibres. *Nat Photonics* **7**, 354–362 (2013).

7. Kress, B. C. Optical architectures for augmented-, virtual-, and mixed-reality headsets. in (Society of Photo-Optical Instrumentation Engineers, 2020).

8. Rademacher, G. *et al.* *.53 Peta-bit/s C-Band Transmission in a 55-Mode Fiber*.

9. Choi, Y. *et al.* Scanner-free and wide-field endoscopic imaging by using a single multimode optical fiber. *Phys Rev Lett* **109**, 203901 (2012).

10. Čižmár, T. & Dholakia, K. Exploiting multimode waveguides for pure fibre-based imaging. *Nat Commun* **3**, (2012).

11. Popoff, S. M. *et al.* Measuring the transmission matrix in optics: an approach to the study and control of light propagation in disordered media. *Phys Rev Lett* **104**, 100601 (2010).

12. Loterie, D. *et al.* Digital confocal microscopy through a multimode fiber. *Opt Express* **23**, 23845–23858 (2015).

13. Li, S. *et al.* Compressively sampling the optical transmission matrix of a multimode fibre. *Light Sci Appl* **10**, (2021).

14. Caramazza, P., Moran, O., Murray-Smith, R. & Faccio, D. Transmission of natural scene images through a multimode fibre. *Nat Commun* **10**, (2019).

15. Borhani, N., Kakkava, E., Moser, C. & Psaltis, D. Learning to see through multimode fibers. *Optica* **5**, 960 (2018).

16. Rahmani, B., Loterie, D., Konstantinou, G., Psaltis, D. & Moser, C. Multimode optical fiber transmission with a deep learning network. *Light Sci Appl* **7**, (2018).

17. Zhu, C. *et al.* Image reconstruction through a multimode fiber with a simple neural network architecture. *Sci Rep* **11**, (2021).

18. Rahmani, B. *et al.* Actor neural networks for the robust control of partially measured nonlinear systems showcased for image propagation through diffuse media. *Nat Mach Intell* **2**, 403–410 (2020).

19. Randel, S., Winzer, P. J., Montoliu, M. & Ryf, R. *Complexity Analysis of Adaptive Frequency-Domain Equalization for MIMO-SDM Transmission*.

20. Arik, S. Ö., Kahn, J. M. & Ho, K. P. MIMO signal processing for mode-division multiplexing: An overview of channel models and signal processing architectures. *IEEE Signal Process Mag* **31**, 25–34 (2014).

21. Fan, P. *et al.* Learning enabled continuous transmission of spatially distributed information through multimode fibers. *Laser Photon Rev* **15**, 2000348 (2021).

22. Caramazza, P., Moran, O., Murray-Smith, R. & Faccio, D. Transmission of natural scene images through a multimode fibre. *Nat Commun* **10**, (2019).

23. Tobin, J. *et al.* Domain Randomization for Transferring Deep Neural Networks from Simulation to the Real World. (2017).

24. Zhao, W., Queralta, J. P. & Westerlund, T. Sim-to-Real Transfer in Deep Reinforcement Learning for Robotics: A Survey. in *2020 IEEE Symposium Series on Computational Intelligence, SSCI 2020* 737–744 (Institute of Electrical and Electronics Engineers Inc., 2020). doi:10.1109/SSCI47803.2020.9308468.

25. Ju, H., Juan, R., Gomez, R., Nakamura, K. & Li, G. Transferring policy of deep reinforcement learning from simulation to reality for robotics. *Nature Machine Intelligence* vol. 4 1077–1087 Preprint at https://doi.org/10.1038/s42256-022-00573-6 (2022).

26. Poole, B., Ozair, S., Oord, A. van den, Alemi, A. A. & Tucker, G. On Variational Bounds of Mutual Information. (2019).

27. Prokudin, S. *Deep Directional Statistics: Pose Estimation with Uncertainty Quantification*.

28. MacKay, D. J. C. & Mac Kay, D. J. C. *Information theory, inference and learning algorithms*. (Cambridge university press, 2003).

29. Zhao, T., Ourselin, S., Vercauteren, T. & Xia, W. Seeing through multimode fibers with real-valued intensity transmission matrices. *Opt Express* **28**, 20978 (2020).

30. Papadopoulos, I. N., Farahi, S., Moser, C. & Psaltis, D. High-resolution, lensless endoscope based on digital scanning through a multimode optical fiber. *Biomed Opt Express* **4**, 260–270 (2013).

31. Baek, Y., Lee, K., Shin, S. & Park, Y. Kramers–Kronig holographic imaging for high-space-bandwidth product. *Optica* **6**, 45–51 (2019).

32. Urness, A. C., Wilson, W. L. & Ayres, M. R. Homodyne detection of holographic memory systems. in *Optical Data Storage 2014* vol. 9201 92010Y (SPIE, 2014).

33. Li, T.-M., Aittala, M., Durand, F. & Lehtinen, J. Differentiable monte carlo ray tracing through edge sampling. *ACM Transactions on Graphics (TOG)* **37**, 1–11 (2018).

34. Nimier-David, M., Vicini, D., Zeltner, T. & Jakob, W. Mitsuba 2: A retargetable forward and inverse renderer. *ACM Transactions on Graphics (TOG)* **38**, 1–17 (2019).

35. Sun, Q., Wang, C., Qiang, F., Xiong, D. & Wolfgang, H. End-to-end complex lens design with differentiable ray tracing. *ACM Trans. Graph* **40**, 1–13 (2021).

36. *https://1v9a14.a2cdn1.secureserver.net/wp-content/uploads/2022/03/SLM-HSP-UHSP-1024-x-1024-Data-Sheet-Rev.0822.pdf*. (2022).

37. *https://www.seagate.com/content/dam/seagate/migrated-assets/www-content/datasheets/pdfs/exos-x20-channel-DS2080-2111GB-en_GB.pdf*.

38. Vaswani, A. *et al.* Attention is all you need. *Adv Neural Inf Process Syst* **30**, (2017).

39. Paszke, A. *et al.* Pytorch: An imperative style, high-performance deep learning library. *arXiv preprint arXiv:1912.01703* (2019).

40. Ronneberger, O., Fischer, P. & Brox, T. U-net: Convolutional networks for biomedical image segmentation. in *International Conference on Medical image computing and computer-assisted intervention* 234–241 (2015).

# ARTICLE

Supplementary information

# Efficient data transport over multimode light-pipes with Megapixel images using differentiable ray tracing and Machine-learning

Joowon Lim [1]*, Jannes Gladrow [1]*, Douglas Kelly [1], Greg O'Shea[1], Govert Verkes [1†], Ioan Stefanovici [1], Sebastian Nowozin [1†], and Benn Thomsen [1]

* Contributed equally to this work

† Work done at Microsoft Research

[1] Microsoft Research, 21 Station Road, Cambridge CB1 2FB, United Kingdom.

Corresponding author: Jannes Gladrow, jannes.gladrow@microsoft.com

1. **Light-pipe cross section comparison: Square vs Circular shapes**

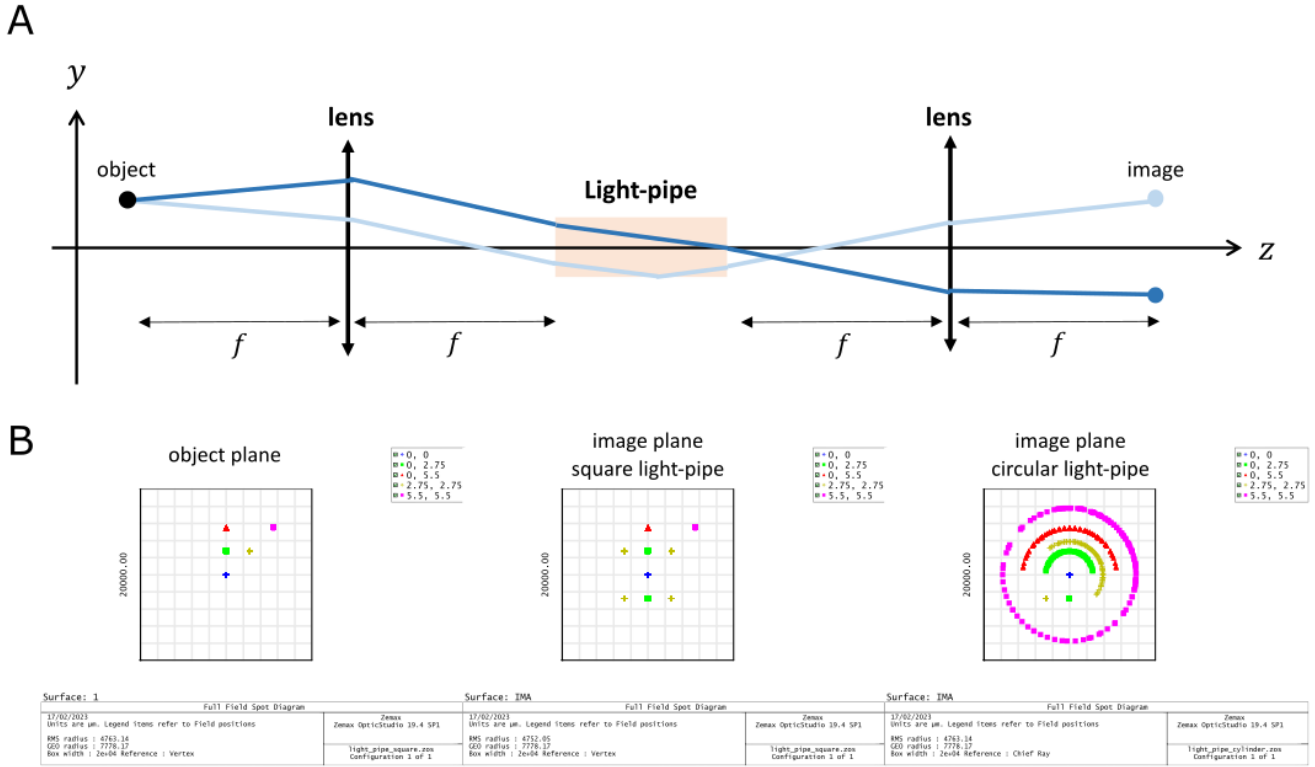

Figure S1. Information smearing for square and circular light-pipe cross-sections. **A** schematic description of ray tracing through a light-pipe. Two rays starting from the same location on the object plane reach different points on the image plane depending on the propagation angle. **B** Spot diagram of rays at the object and image planes for square and circular cross-sections. The numbers in the legends represent $x$ and $y$ values in millimeters, and the size of spot diagram is $2\ mm \times 2\ mm$.

Fig. S1 shows a schematic diagram of rays propagating through a light-pipe. As shown in Fig. S1A, rays emanating from a point form images at two different locations depending on the propagation direction. This figure depicts the ray tracing only in 2D (y-z), but we can expand this intuition to 3D, however, we then need to consider the shape of light-pipe cross-section and its effect on the distribution of rays. We performed the ray tracing analysis on two different cross-section shapes, square and circular shapes, assuming the lenses are perfect paraxial lenses using Zemax OpticsStudio (22.1.2). We set the numerical aperture of rays from the object as 0.0145 (which corresponds to the case of 2x2 symbol size in the main paper) and varied the location of the object as shown in Fig. S1B. In the case of the square light-pipe, when the position of object is close to the optical axis, all the rays reach the single point on the image plane (blue cross), and they start to diverge leading to two different locations as we move along the y-axis (green square). However, after a certain point, they start to converge to a single point (red triangle) again, resulting in a perfect imaging system (except for the flip). If we move the object along both x and y axes, we can see that rays diverge to 4 points (yellow cross) but they go back to a single point again (purple square).

For the circular cross-section, when the object position is close to the optical axis, all the rays also converge in a single point on the image plane (blue cross). However, unlike the square case, as we move the object further from the optical axis, you can see one spot formed from rays that are not reflected within the light-pipe and the radial divergent pattern of rays formed from reflected rays (green square). This is in contrast to the previous case where we saw only two spots. As we move further to a point where all the rays are reflected, all rays are now radially smeared (red triangle), which is in stark contrast to the previous case where all rays converge to a single point. We can see the same tendency when we move the object diagonally (yellow cross and purple square).

Therefore, you can see that the information of a single symbol will be smeared over multiple locations as the size of object increases and the degree of smearing dramatically depends on the shape of light-pipe cross-section. Given a fixed noise level, this inherently limits the capacity of channel, as the high-entropy data signal is smeared spatially and summed, which reduces the certainty about original symbol when corrupted by noise. Therefore, it is important to keep our optical system as close as possible to an imaging system, which we can achieve by using a light-pipe with a square cross-section rather than a commonly used circular cross-section fiber.

## 2. Light-pipe imaging condition

To preserve full information, the width and NA of light-pipe should satisfy the following conditions. To be specific, when the pixelated discrete data is presented on spatial light modulator (SLM) whose pitch is $p$ over the length of $L$ as shown in Fig. S2, we need to make sure to capture the first order of it, $\frac{\lambda f}{p}$, to ensure the retrieval of the original data where $\lambda$ is the wavelength and $f$ is the focal length of the lens. In other words, the width of light-pipe, $W$, should be bigger than $\frac{\lambda f}{p}$. In addition, the numerical aperture of light-pipe should be bigger than $\sin\left(\tan^{-1}\left(\frac{L}{2f}\right)\right) \approx \frac{L}{2f}$ where the approximation comes under the paraxial approximation.

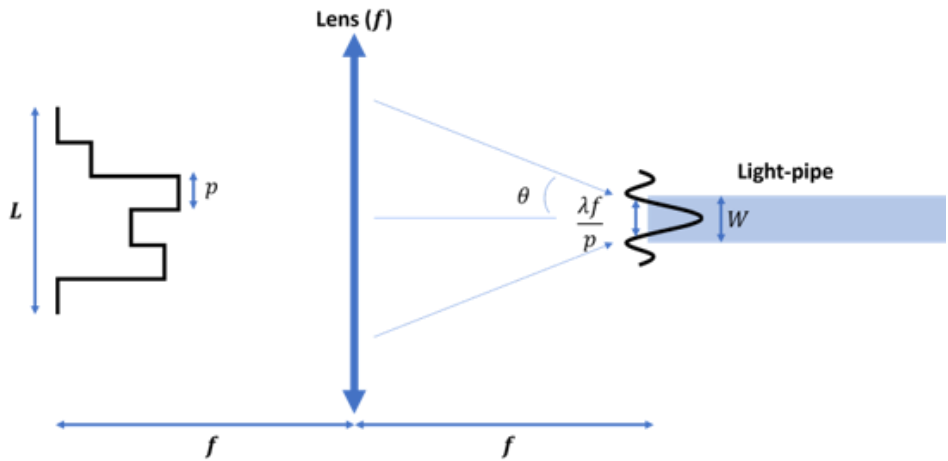

Figure S2. Schematic description of the information preserving condition of focused light onto light-pipe.

## 3. Decoder architecture

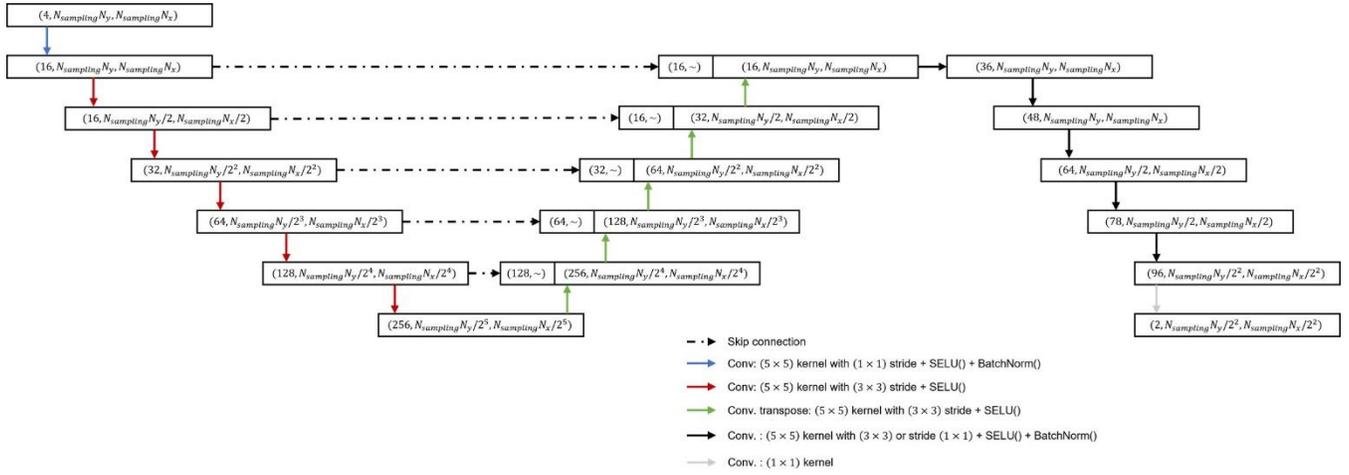

Figure S3. Schematic description of the decoder structure. In each block, the first number represents the channel number. The second and the third numbers represent the pixel numbers of image. For the skip-connection block, we only represent the channel number and omit the image pixels. $N_{sampling}$ is a positive integer value to decide the sampling factor given symbol numbers along $x$ and $y$ axes, $N_x$ and $N_y$, respectively.

The decoder network architecture is displayed in Fig. S3. The input has 4 channels: two sampled camera images and two positional encoding images ($x$ and $y$ 2D grid images, shown in Fig. 4D and Fig. S6A, C). The input image size is $(N_{sampling}N_x, N_{sampling}N_y)$ where $N_{sampling}$ is an positive integer value to decide the sampling factor given symbol numbers along $x$ and $y$ axes, $N_x$ and $N_y$, respectively. Depending on $N_{sampling}$, we adjust the stride values of convolution layers after the up-sampling blocks. In Fig. S3, the stride values for the convolution layers marked in black arrows are (1, 1, 3, 1, 3), which results in a 4-times down sampling along each axis. The architecture in Fig. S3 is thus for $N_{sampling} = 4$. For $N_{sampling} = 8$ and 16, we change the stride values to (1, 1, 3, 3, 3) and (1, 3, 3, 3, 3), respectively.

## 4. Parametrization of Network Outputs

As explained in the main text, our U-Net decoder predicts a phase-angle $\mu_{ij}$ and concentration $\kappa_{ij}$ for each symbol $(i,j)$. Letting the model directly output $\mu_{ij}$ and $\kappa_{ij}$ separately causes unwrapping problems at the $0 - 2\pi$ boundary for $\mu_{ij}$. It also requires neurons in the last layer to specialize to either angles or concentrations, leading to worsened performance and even numeric instability. We therefore let the model output a 2D vector $\vec{p}_{ij} = (p_{ij}^x, p_{ij}^y)$ for each symbol. The phase-angle is represented by the polar angle of $\vec{p}_{ij}$, $\mu_{ij} = \arctan2\left(p_{ij}^y / p_{ij}^x\right)$ while the norm of the vector controls the concentration, i.e. $\kappa_{ij} = \|\vec{p}_{ij}\|$.

## 5. Differentiable Bessel Functions

In order to implement the von-Mises log-likelihood loss, we need a differentiable implementation of $\log\left(I_m(\exp(x))\right)$ for $m = 0, 1$ in pytorch. We use the following approximations[1]:

$\log(I_0(e^x)) \approx \log(1 + 3.5156229\ t(x)^2 + 3.0899424\ t(x)^4 + 1.2067492\ t(x)^6 + 0.2659732\ t(x)^8 + 0.0360768\ t(x)^{10} + 0.0045813\ t(x)^{12})$ for $x < \log(3.75)$ with $t(x) = e^x/3.75$ and otherwise $\log(I_0(e^x)) \approx e^x - 0.5\ x + \log(0.39894228 + 0.01328592\ t(x)^{-1} + 0.00225319\ t(x)^{-2} - 0.00157565\ t(x)^{-3} + 0.00916281\ t(x)^{-4} - 0.02057706\ t(x)^{-5} + 0.02635537\ t(x)^{-6} - 0.01647633\ t(x)^{-7} + 0.00392377\ t(x)^{-8})$.

For m=1, we have for $x < \log(3.75)$, $\log(I_1(e^x)) \approx x + \log(0.5 + 0.87890594\ t(x)^2 + 0.51498869\ t(x)^4\ 0.15084934\ t(x)^6 + 0.02658733\ t(x)^8 + 0.00301532\ t(x)^{10} + 0.00032411\ t(x)^{12})$ and otherwise $\log(I_1(e^x)) \approx e^x - 0.5\ x + \log(0.39894228 - 0.03988024\ t(x)^{-1} - 0.00362018\ t(x)^{-2} + 0.00163801\ t(x)^{-3} + 0.01787654\ t(x)^{-7} - 0.00420059\ t(x)^{-8})$.

## 6. Network training

In all cases, we use a batch size of 2, given the already high degree of weight-sharing over all symbols on a data page in our convolutional model. An additional, practical consideration here is the high memory load due to the large size of the input camera images. We use two different learning rates; one for the decoder parameters (see Table 1) and the other for the affine transform parameters (2e-4, please refer to Fig. S6 and the corresponding content regarding the affine transform). For both learning rates, we linearly warm up the learning rates to the target values (as noted in Table 1) for an epoch number corresponding to 2% of total epoch number. For the decoder learning rate, we damp it by the learning rate decay rate (see Table 1) at epoch numbers corresponding to 80% and 90% of total epoch numbers. In the case of the parallelogram-angle-fitting scenario (Fig. 4 in the main text), we use a different learning rate (1e-4 for experimental data and 0.0 for simulation data since the simulations are based on a perfect square cross-section light-pipe) to update the parallelogram angle. We use the Adam optimizer[2] in all cases. When updating the decoder model parameters, we clip all gradients whose norm is larger than 1.0. For faster convergence, we initialize the diagonal values of the affine transform (scaling parameters). These values are set to approximately match the ratio between the field sizes of camera and SLM in the $x$ and $y$ axes, respectively. These values are (0.98, 0.95). Also, for simulation data, when we use the ray-based localizer, we assume perfect paraxial lenses while we model actual compound lenses for experimental data.

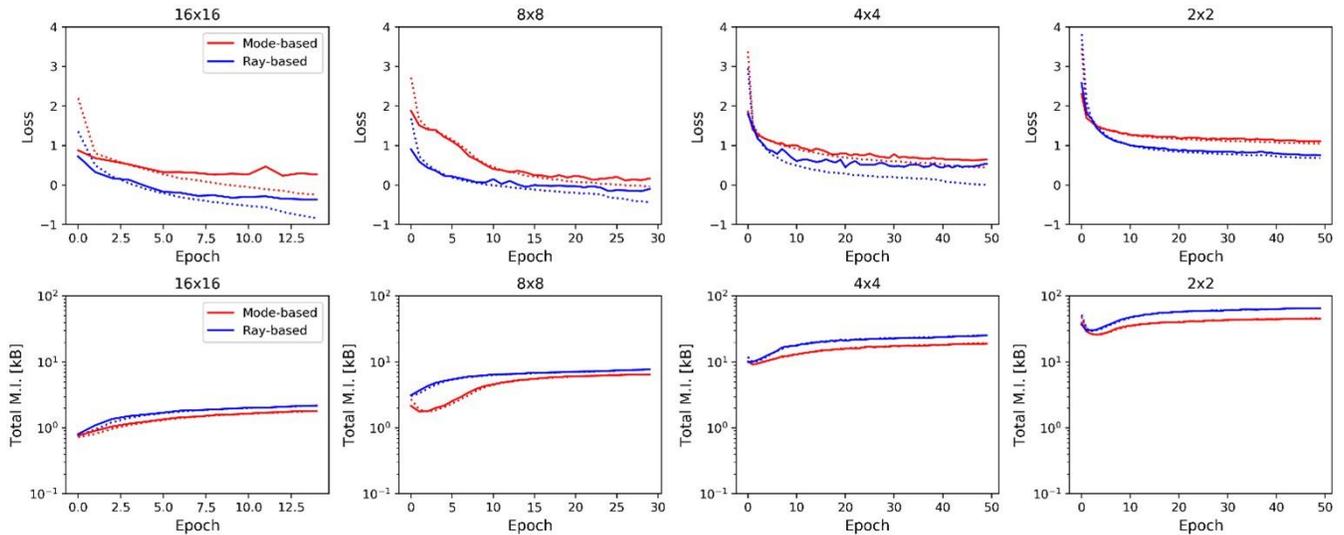

Figure S4. Train/validation loss (upper row) and total MI (lower row) curves for experiment data with various symbol sizes, and each symbol size is written on top of the graphs. The red and blue lines represent the mode-based and ray-based localizers, respectively. The dotted lines are train curves, and the solid lines are validation curves.

Figure S4 shows loss and total mutual information (MI) curves on train and valid datasets during training for different symbol sizes. As expected from the final performance shown in the main text, we can see that the ray-based localizer always shows higher MI values than the ones of the mode-based localizer. For smaller symbol sizes (8x8 and 16x16), we reduced the epoch numbers to 15 (16x16) and 30 (8x8), as they converge fast and start to be overfitted. Also, with the decrease of symbol size, the loss values become higher overall, which explain it becomes hard to extract information from smaller number of pixels.

Table 1. Training hyperparameters for experimental data.

|  | Symbol size | Epoch number | Decoder learning rate | Learning rate decay rate | Stride lists | $N_{sampling}$ |
|---|---|---|---|---|---|---|
| Mode-based | 2 x 2 | 50 | 1e-3 | 0.5 | [1, 1, 3, 1, 3] | 4 |
|  | 4 x 4 | 50 | 1e-3 | 0.5 | [1, 1, 3, 1, 3] | 4 |
|  | 8 x 8 | 30 | 1e-3 | 1.0 | [1, 1, 3, 3, 3] | 8 |
|  | 16 x 16 | 15 | 1e-3 | 1.0 | [1, 3, 3, 3, 3] | 16 |
| Ray-based | 2 x 2 | 50 | 1e-3 | 0.5 | [1, 1, 3, 1, 3] | 4 |
|  | 4 x 4 | 50 | 1e-3 | 0.5 | [1, 1, 3, 1, 3] | 4 |
|  | 8 x 8 | 30 | 1e-3 | 1.0 | [1, 1, 3, 3, 3] | 8 |
|  | 16 x 16 | 15 | 1e-3 | 1.0 | [1, 3, 3, 3, 3] | 16 |

Table 2. Training hyperparameters for simulation data.

|  | light-pipe length | Epoch number | Decoder learning rate | Learning rate decay rate | Stride lists | $N_{sampling}$ |
|---|---|---|---|---|---|---|
| Mode-based | All cases | 50 | 1e-3 | 0.5 | [1, 1, 3, 1, 3] | 4 |
| Ray-based | All cases | 50 | 1e-3 | 0.5 | [1, 1, 3, 1, 3] | 4 |

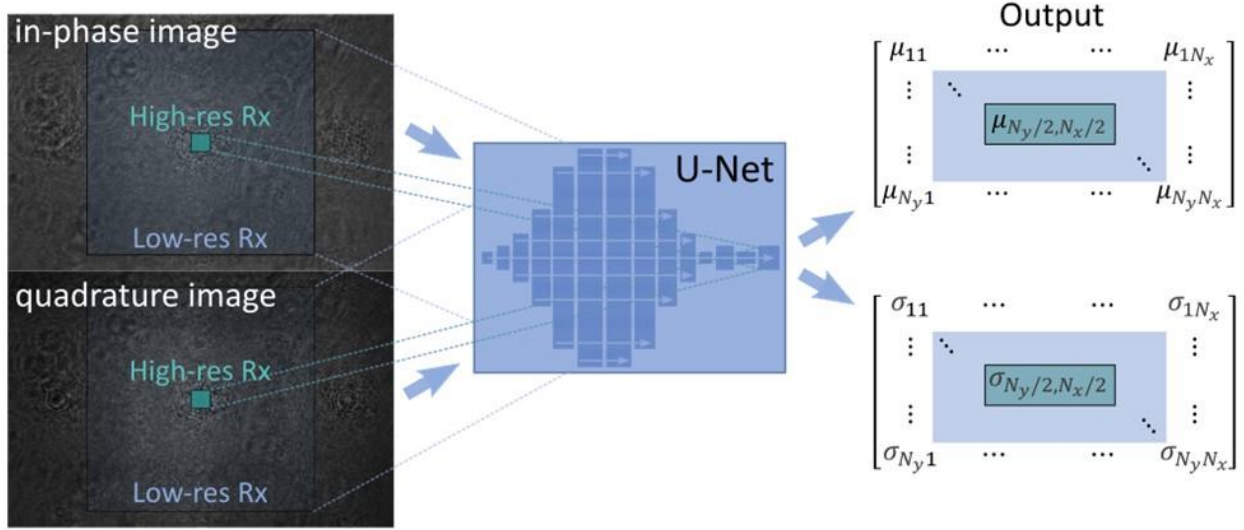

Figure S5. Receptive fields of the U-Net architecture. The U-Net processes the image on different spatial scales. The layers operating at the highest resolution are mainly responsible for decoding individual symbols but can consider information from the surrounding provided by coarser layers. The receptive field of the bottleneck 'U' part is denoted as 'Low-res' while the receptive field of the main branch of the network is denoted as 'High-res'.

## 7. Receptive Field of our U-Net

As discussed in the main text and in Fig. S5, the need for special localizers arises because the convolution-based decoder, while efficient, has only a limited receptive field per symbol and a static set of convolutional kernels. In fact, the particular 'U' part of the U-Net structure endows the network with receptive fields on different resolutions: A top layer of the network 'sees' inputs on the highest resolution, while the down- and up-sampling layers (the 'U-part') of the network see the inputs on an exponentially coarser resolution, the deeper the layer. The deepest layers in the down-sampling part of the model can thus incorporate information on global illumination or global shifts of the grid. However, these layers cannot be expected to decode individual symbols, especially if the number of filters (channel number) is not high enough to route information from every symbol on the input side to every symbol to the output side. Combinatorially, the required filter number to achieve this may be prohibitively high and we never achieved this regime. In Table 3, we list the receptive fields over the different branches of the U-Net.

Table 3. Receptive fields of different layers in the decoder. The down- and up-sampling layers are skipped here since they contribute information on a different coarseness as discussed in the text.

| Layer | Kernel size | Stride | Receptive Field (Size in symbols at $N_{sampling} = 4$) |
|---|---|---|---|
| 1 | 5 | 1 | 1.25 |
| 2 | 5 | 1 | 2.25 |
| 3 | 5 | 1 | 3.25 |
| 4 | 5 | 2 | 7.25 |
| 5 | 5 | 1 | 8.25 |
| 6 | 5 | 2 | 17.25 |
| 7 | 1 | 1 | 17.25 |

## 8. Detailed Description of Localizer Modules

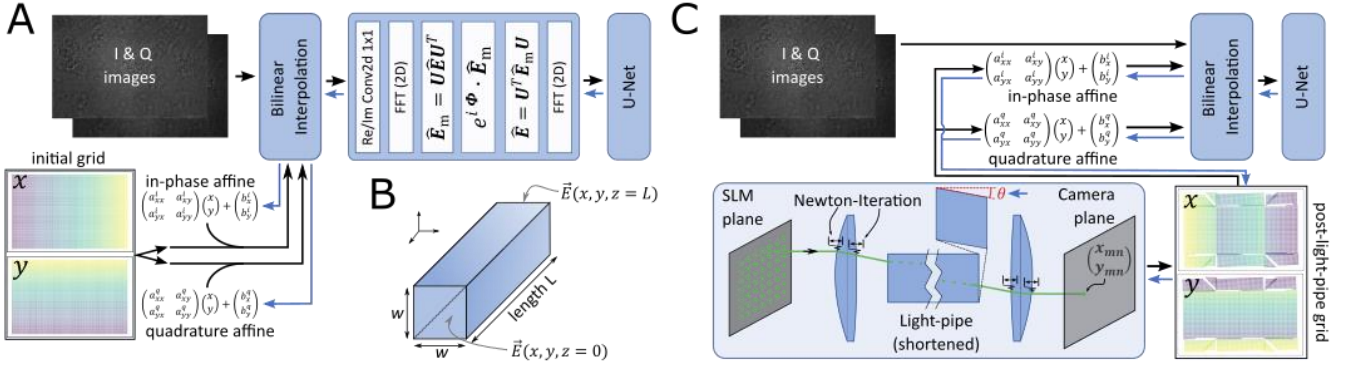

Figure S6. Details of the localizer modules. Forward operations are indicated by black arrows, while blue arrows indicate the flow of backward passes that build the gradient during training. **A** Mode-based localization pipeline as described in the text. **B** Description and definitions of variables of the light-pipe. **C** Ray-based localization pipeline. The localization in this pipeline is entirely handled by the sampling grid, which is obtained by ray-tracing through a precise digital twin of the setup. The light-pipe-parallelogram angle is shown in red to indicate that this variable receives a gradient and can be trained.

## 9. Mode-based localizer

The mode-based pipeline is shown in Fig. S6A. First, since our camera images are giga-pixel images, we down-sample the camera images to $(N_{sampling}N_x, N_{sampling}N_y)$ where $N_x$ and $N_y$ represent symbol numbers (not pixel numbers) along $x$ and $y$ axes on SLM and $N_{sampling}$ is the sampling rate we can control. To decide the sampling grid points on the camera images, we roughly estimate the start and end points along both $x$ and $y$ axes and perform equi-distance sampling, which give us $x$ and $y$ grid points. Since we use two cameras, the captured images from them should be well registered to each other in order to extract the electric field information. So, we apply affine transform independently on two independent sets of grids ($x$ and $y$ grids for I camera and another set of $x$ and $y$ grids for Q camera) to correct potential misalignment in real experiment. The sampled images are fed into the mode-based waveguide inversion, which again, is differentiable. The output of the mode-based inversion is then inserted into the U-Net decoder.

Here, we describe the inversion step in detail. The propagation of TE-modes in a square-shaped light-pipe is a classic problem. The propagator of the electrical field can be obtained from solving Helmholtz' Equation $\Delta E(x,y,z) + \varepsilon_r(x,y,z)k^2 E(x,y,z) = 0$ where $E(x,y,z)$ is the electrical field, and $\varepsilon_r(x,y,z)$ and $k = 2\pi/\lambda$ are the relative electric permittivity value (which is square of refractive index value) and free space wave-number respectively. The equation comes with boundary conditions; here we assume absorbing boundaries as an approximation, i.e., $\vec{E}(x,y,z) = 0$ on the boundaries. Frist, we describe the forward propagation through light-pipe. As shown in Fig. S6B, given a square light-pipe whose width (height) and length are $w$ and $L$, respectively, we can solve for the transversal field at the distal facet of light-pipe, $E_z(x,y,z=L) = \sum_{n=1}^{\infty}\sum_{m=1}^{\infty} E_0^{(m,n)} \sin\frac{m\pi}{w}x \sin\frac{n\pi}{w}y \, e^{ik_z^{(m,n)}L}$ with $k_z^{(m,n)} = \sqrt{(n_{lp}k)^2 - \left(\frac{m\pi}{w}\right)^2 - \left(\frac{n\pi}{w}\right)^2}$ and $E_0^{(n,m)}$ denoting the mode coefficients at the frontal facet of light-pipe and $n_{lp}$ is the refractive index value of light-pipe. Each mode coefficient, $E_0^{(m,n)}$, can be calculated by projecting the electric field at the frontal facet, $E_z(x,y,z=0)$, onto the corresponding sinusoidal mode base, $\sin\frac{m\pi}{w}x \sin\frac{n\pi}{w}y$. For a discretized field, $E_z(x_i, y_j, z=0)$ where $i = 1..N_x$ and $j = 1..N_y$, this can be efficiently implemented using matrix multiplication. In other words, $E_0^{(m,n)}$ is the $(m,n)$ component of the resulting matrix of $U_y^T E_z(x_i, y_j, z=0) U_x$, where $U_x = \left[\sin\left(\frac{0\vec{x}\pi}{w}\right), \sin\left(\frac{1\vec{x}\pi}{w}\right), \ldots, \sin\left(\frac{M\vec{x}\pi}{w}\right)\right]$ and $U_y = \left[\sin\left(\frac{0\vec{y}\pi}{w}\right), \sin\left(\frac{1\vec{y}\pi}{w}\right), \ldots, \sin\left(\frac{N\vec{y}\pi}{w}\right)\right]$ given $\vec{x} = [x_0, x_1, \ldots, x_{N_x-1}]^T$ and $\vec{y} = [y_0, y_1, \ldots, y_{N_y-1}]^T$. The mode coefficient at the distal facet can be calculated by multiplying $E_0^{(m,n)}$ with the corresponding modal dispersion factor, $e^{ik_z^{(m,n)}L}$, resulting in $E_L^{(m,n)}$. By denoting the mode coefficient matrix at the distal facet as $E_L$ and these coefficients can be transformed back to the image domain by calculating $E_z(x_i, y_j, z=L) = U_y E_L U_x^T$. The inversion step can be easily deduced from the forward step; we just need to use the conjugated value for the modal dispersion term, and the other processes stay unchanged. These operations can be implemented in Pytorch[3], including the 2f-launch and 2f-delaunch lenses, which are modeled by fast Fourier transform. Since these mode transformations are differentiable due to Pytorch's ability to compute derivates automatically, we add a 1x1-conv2d layer in front of the mode-based localized which acts separately on in-phase and quadrature images. For this convolution layer, we initialize weight values to 1 and bias values to 0. This allows the model to automatically account for any global phase offset or any difference in image-wide magnitudes between in-phase and quadrature images acquired from two different cameras. Downstream from the camera registration (which we explain below) and aforementioned 1x1-conv2d layer, we combine in-phase and quadrature tensors into a complex-valued 2D array.

### 10. Ray-based localizer

The ray-based pipeline is shown in Fig. S6C. An initial grid of points (one set of $x$ and $y$ grids) on the SLM plane is used as origin of rays in our ray-based localizer. The rays propagate through the digital twin of the setup and get refracted at the first lens surface. We model our lenses as thick lenses which requires analytical descriptions of the lens surface shapes. Lens profiles are usually expressed in terms of the sag which quantifies the respective distance of the surface from the frontal or back plane of the lens. We define the sag as a function of the transversal distance from the optical axis, $r$: $\mathrm{sag}(r) = \frac{\kappa r^2}{1+\sqrt{1-Ar^2}}$ where $A = (1+c)\kappa^2$ with $\kappa, c$ denoting the curvature and the conic coefficient, respectively. The gradient of the sag with respect to a coordinate position $(x,y,z)$ can then be analytically derived which is sufficient to implement a Newton iteration for the ray-surface intersection[4]. For our virtual lens, we use the curvatures and refractive index provided by the vendor (see experimental method section) such that we have a close analogue of the lens used in the experimental setup. For refraction, we apply Snell's law to each ray. The light-pipe is implemented as a refractive cuboid with dimensions $3.56 \times 3.56 \times 75 \text{ mm}^3$. The refractive index is $n_{lp} = 1.461$. Rays are propagated from the lens onto the frontal facet of the light-pipe where they refract. Inside the light-pipe, rays are reflected off the walls if they obey the total internal refraction angle. On the distal facet of the light-pipe, rays refract back into air and propagate onto the next lens where they get imaged onto the camera plane. In the ray-tracer, we ignore the optical elements of the homodyne detector (see Fig. 1 in the main text) and propagate the rays onto a single camera since we only need to extract positions of rays. Two independent affine transformations (for I&Q) are used again to ensure co-registration of the two cameras. The final grids are then used to sample the images using bilinear interpolation. The resulting image has a restored locality such that we can feed it into the U-Net decoder.

### 11. Mode analysis for a parallelogram light-pipe

Here, we try to find an analytical mode formulation for a parallelogram cross-section light-pipe and demonstrate how it differs from the one of a square cross-section light-pipe. First, we start from the same governing equation, Helmholtz' Equation: $\Delta \vec{E}(x,y,z) + \varepsilon_r(x,y,z)k^2\vec{E}(x,y,z) = 0$, with the absorbing boundary condition ($E_{0,z}(x,y) = 0$ when $x = w$ or $y = w$ where $w$ is the width/height of light-pipe.). We can represent a mode with propagation speed, $\beta$, as follows: $E(x,y,z) = E_{0,z}(x,y)e^{i\beta z}$ where $E_{0,z}(x,y)$ represents the mode profile in the transverse $(x,y)$ plane. Plugging this in Helmholtz' Equation, it results in $\left(\frac{\partial^2}{\partial x^2} + \frac{\partial^2}{\partial y^2}\right)E_{0,z}(x,y) + k_c^2 E_{0,z}(x,y) = 0$, where $k_c^2 = \varepsilon_r(x,y,z)k^2 - \beta^2$. In deriving the mode bases for a square cross-section light-pipe, we represent the solution of mode as $E_{0,z}(x,y) = X(x)Y(y)$, which is the multiplication of two independent functions ($X(x)$ and $Y(y)$) along $x$ and $y$ axes. Therefore, Helmholtz' Equation can be simplified as follows: $\frac{1}{X(x)}\frac{\partial^2}{\partial x^2}X(x) + \frac{1}{Y(y)}\frac{\partial^2}{\partial y^2}Y(y) + k_c^2 = 0$. Using the absorbing boundary condition, we can get $X(x) = \sin\left(\frac{m\pi}{w}x\right)$ and $Y(y) = \sin\left(\frac{n\pi}{w}y\right)$ and the modes are represented as the multiplication of these functions for various integer numbers ($m$ and $n$). We will apply the same logic for a parallelogram light-pipe, but we will represent the solution in $y'$ axis $E_{0,z}(x,y) = X(x)Y'(y')$, where $y' = x\sin\theta + y\cos\theta$ given a parallelogram angle $\theta$. By plugging it into Helmholtz' Equation, it results in $\frac{1}{X(x)}\frac{\partial^2}{\partial x^2}X(x) + \frac{1}{Y'(y')}\frac{\partial^2}{\partial y'^2}Y'(y') - \frac{4\sin\theta}{(1+\sin^2\theta)X(x)Y'(y')}\frac{\partial}{\partial x}X(x)\frac{\partial}{\partial y'}Y'(y') + k_c^2 = 0$. In this case, we have the term which depends both on $X(x)$ and $Y'(y')$, which disappear when the parallelogram angle becomes 0. In other words, for a parallelogram light-pipe, we are not able to analytically model the mode as the multiplication of sinusoidal functions along $x$ and $y'$.

### 12. In-phase and quadrature image formation

The homodyne detector shown in Fig. 1A is configured such that the two cameras, when the optical path is unfolded, are coincident along the optical axis (by convention, $z$). Thus, the captured in phase and quadrature images can be used to generate the spatially- ($xy$-) varying real and imaginary parts of the field respectively at the single notional detection plane. To demonstrate this, we set out a simplified explanation of the formation of images in the homodyne detector where, for brevity, each of the optical components described (waveplates, beam splitter, polarizing beam splitter, camera) is assumed to perform ideally. In Fig. S7, we present how the polarization states change while the beams travel.

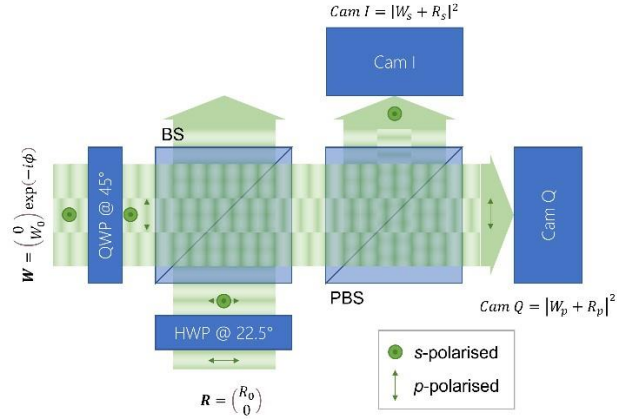

Figure S7. Illustration of the homodyne detector showing the polarization states at different positions in the system.

Consider the fields of the signal (W) and reference (R) beams before entering the homodyne detector. Omitting $z$-dependent phase for clarity, we can capture each field thus:

$$\boldsymbol{W} = \begin{pmatrix} 0 \\ W_0 \end{pmatrix} \exp(-i\phi)$$
$$\boldsymbol{R} = \begin{pmatrix} R_0 \\ 0 \end{pmatrix}, \tag{1}$$

where the amplitude vector represents components in the ($p$, $s$)-polarizations for each beam, and $\phi = \phi_{SLM}(x, y)$ is the $xy$-varying phase retardation imparted on the signal beam by the SLM.

The actions of the quarter wave and half wave plates on the signal and reference beams respectively result in the beams at the entrance to the beam splitter being of the form:

$$\boldsymbol{W} = \frac{1}{\sqrt{2}} W_0 \begin{pmatrix} \exp\left(i(-\phi + \pi/2)\right) \\ \exp(-i\phi) \end{pmatrix}$$
$$\boldsymbol{R} = \frac{1}{\sqrt{2}} R_0 \begin{pmatrix} 1 \\ 1 \end{pmatrix}. \tag{2}$$

In the implementation shown in Fig. 1A, half of the light from each beam is discarded at the beam splitter; the other halves, now travelling coaxially, are routed according to polarization at the polarizing beam splitter such that the *s*-polarization of each beam is reflected, and the *p*-polarization transmitted. Then, the fields in each polarization can be written:

$$S = \frac{1}{2\sqrt{2}} [W_0 \exp(-i\phi) + R_0]$$
$$P = \frac{1}{2\sqrt{2}} \left[W_0 \exp\left(i(-\phi + \pi/2)\right) + R_0\right] \tag{3}$$

The cameras in the s- and p-polarization paths "see" the *intensity* of the field incident upon them:

$$|S|^2 = SS^* = \frac{1}{8}(W_0^2 + R_0^2 + 2W_0 R_0 \cos\phi)$$
$$|P|^2 = PP^* = \frac{1}{8}(W_0^2 + R_0^2 + 2W_0 R_0 \sin\phi).$$
(4)

Thus, following DC-subtraction and renormalization (processing, denoted by subscript *processed*) we can extract the in-phase ($W_0 \cos\phi$) and quadrature ($W_0 \sin\phi$) parts of the $xy$-varying field, and so infer the phase (and amplitude) encoded at the transmitter:

$$Field = I + iQ = |S|^2_{processed} + i|P|^2_{processed}.$$
(5)

## 13. Setup Stability

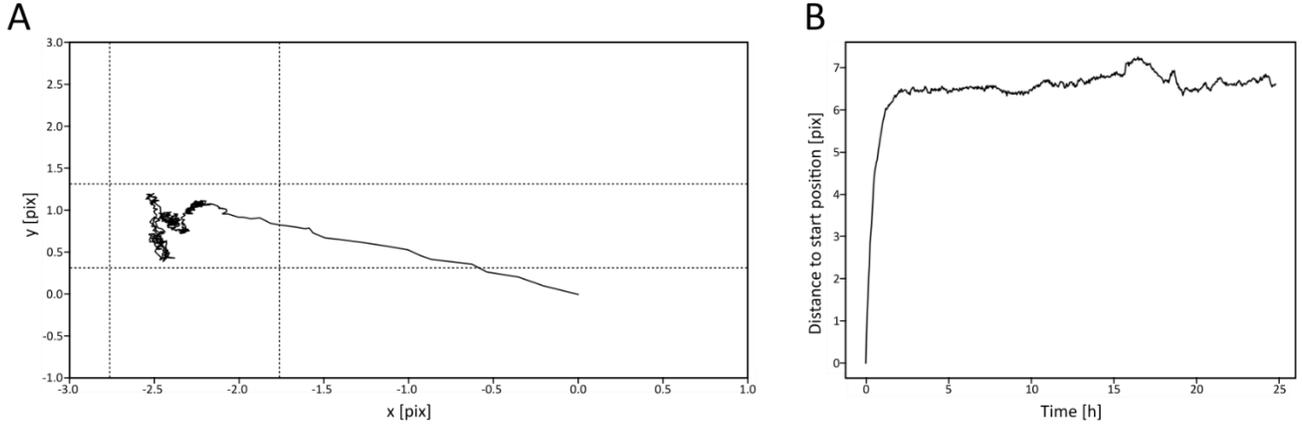

Figure S8. Drift measurements of our setup over 24h. **A** An average drift of crosshair localizations over time with respect to the first measurement. Dashed lines indicate a box of 1x1 pixel side length. **B** Average Euclidean distance to starting position over time.

To get a sense of the stability of the optical setup in Fig. 1A, we carried out a 24-hour measurement in which we collected camera images every 2.5 minutes of a 6x6 grid of crosshairs located on the SLM in such a way that they would end up in the central, undistorted zone of the waveguided image. In order to investigate the stability of light-pipe, we blocked the reference beam in this experiment to simplify marker localization. We localized the crosshairs by first removing the background by subtracting an image convolved with a 15x15-sized gaussian kernel and then convolving with an image of a crosshair. Afterwards, we computed the element-wise square of the image and used the local peak finder algorithm in the scikit-image python package[5]. We then refined the peak locations using a center-of-mass estimator. The measurements show that the setup settles into a stable state after an initial 2~3-hour long warm-up phase. Afterwards, crosshair centers remain within a single pixel on average indicating excellent stability. A camera pixel has a side length of 2.4 μm. We therefore expect our decoder to only require finetuning after very prolonged periods of time.

**References**

1. Milton Abramowitz and Irene Stegun. *Handbook of Mathematical Functions with Formulas, Graphs, and Mathematical Tables*. (1972).

2. Kingma, D. P. & Ba, J. Adam: A method for stochastic optimization. *arXiv preprint arXiv:1412.6980* (2014).

3. Paszke, A. *et al.* Pytorch: An imperative style, high-performance deep learning library. *arXiv preprint arXiv:1912.01703* (2019).

4. Sun, Q., Wang, C., Qiang, F., Xiong, D. & Wolfgang, H. End-to-end complex lens design with differentiable ray tracing. *ACM Trans. Graph* **40**, 1–13 (2021).

5. Van der Walt, S. *et al.* scikit-image: image processing in Python. *PeerJ* **2**, e453 (2014).